\documentclass{elsart3}
\usepackage{graphicx}
\usepackage{amsmath,amssymb}
\begin{document}

\begin{frontmatter}
\title{Critical slowing down in the geometrically frustrated\\ pyrochlore antiferromagnet~Gd$_2$Ti$_2$O$_7$}
\author[UCR]{D.~E. MacLaughlin\corauthref{MacL}}
\corauth[MacL]{Tel. +1-951-827-5344, Fax +1-951-827-4529, e-mail: macl@physics.ucr.edu.}
\author[UCR]{M.~S. Rose}
\author[UCR]{J.~E. Anderson}
\author[UCR]{Lei Shu}
\author[LANL,JAERI]{R.~H. Heffner}
\author[LANL]{T. Kimura}
\author[LANL,TRIUMF]{G.~D. Morris}
\author[CSULA]{O.~O. Bernal}
\address[UCR]{Department of Physics, University of California, Riverside, California 92521, U.S.A.}
\address[LANL]{Los Alamos National Laboratory, K764, Los Alamos, New Mexico 87545, U.S.A.}
\address[JAERI]{Japan Atomic Energy Research Institute, Tokai, Ibaraki-ken, 319-1195, Japan}
\address[TRIUMF]{TRIUMF, Vancouver, Canada \mbox{V6T 2A3}}
\address[CSULA]{Department of Physics and Astronomy, California State University, Los Angeles, California 90032, U.S.A.}

\begin{abstract}
Longitudinal-field muon spin relaxation experiments have been carried out in the paramagnetic state of single-crystal Gd$_2$Ti$_2$O$_7$ just above the phase transition at $T_{\rm m} = 1.0$~K\@. At high applied fields the exponential relaxation time~$T_1$ is proportional to field, whereas $T_1$ saturates below a crossover field~$B_{\rm c}$ that is $\sim$2.5~T at 1.5~K and decreases as $T_{\rm m}$ is approached. At low fields the relaxation rate increases markedly as the freezing temperature is approached, as expected for critical slowing down of the spin fluctuations, but the increase is suppressed by applied field. This behavior is consistent with the very long autocorrelation function cutoff time implied by the low value of $B_{\rm c}$.
\end{abstract}

\end{frontmatter}

\section{Introduction}
In the pyrochlore titanate~Gd$_2$Ti$_2$O$_7$ the Gd$^{3+}$ local moments occupy corner-shared tetrahedral sites, leading to geometrical frustration, enormous ground-state degeneracy, and extreme sensitivity to mechanisms that break this degeneracy~\cite{Rami94}. Phase transitions are not expected for a pyrochlore lattice of spins coupled by a Heisenberg exchange interaction~\cite{MoCh98,MoCh98b}. But Gd$_2$Ti$_2$O$_7$ exhibits a zero-field magnetic phase transition at $T_{\rm m} = 1.0$~K~\cite{RDGM99}; below this temperature a complex field-temperature phase diagram is found~\cite{RSHK02,PLBP04}. The transition temperature is only slightly dependent on applied field, and the data suggest a multicritical point at $\sim$1.5~T and $\sim$0.86~K\@. Similar behavior emerges from theories based on dipolar interactions between Gd$^{3+}$ moments~\cite{RSHK02,CeSh04}. 

Although M{\"o}ssbauer data suggest a first-order transition at $T_{\rm m}$~\cite{BHOS03}, the specific heat and susceptibility measurements that establish the phase diagram show no evidence for a discontinuous jump or hysteresis at the transition for any field, suggesting that all transitions are of second order. Zero-field muon spin relaxation ($\mu$SR) measurements in the ordered state of Gd$_2$Ti$_2$O$_7$~\cite{YDGM05} indicate that strong dynamic fluctuations persist to low temperatures. 

This paper reports a $\mu$SR study of critical dynamics in the paramagnetic state ($T \gtrsim T_{\rm m}$) of Gd$_2$Ti$_2$O$_7$, which was motivated by the above results and the possibility of strong low-frequency fluctuations resulting from frustration in this compound.

\section{Experimental Results}
Longitudinal-field $\mu$SR experiments were carried out in single-crystal Gd$_2$Ti$_2$O$_7$ at the M15 channel, TRIUMF, Vancouver, Canada. Data were taken at temperatures just above $T_{\rm m}$ in fields~$B$ of up to 4~T applied parallel to the $\langle111\rangle$ direction. Exponential relaxation was observed at all temperatures and fields. 

Figure 1 gives the field dependence of the exponential relaxation time~$T_1$ at three temperatures just above $T_{\rm m}$.
 \begin{figure}[ht]
     \centering
     \includegraphics[width=0.85\columnwidth]{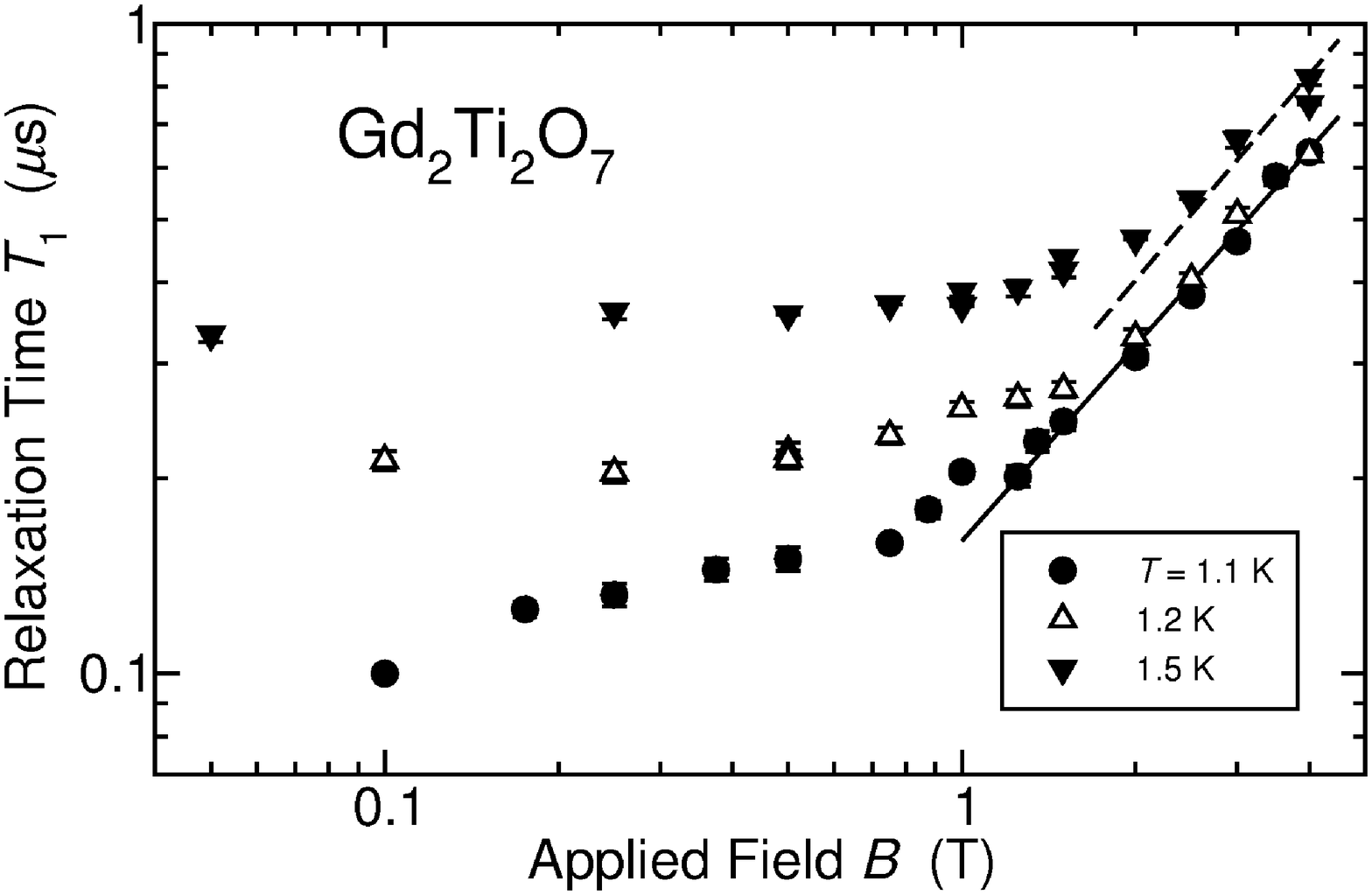}
     \caption{Dependence of the longitudinal-field muon spin relaxation time~$T_1$ on applied field~$B$ in Gd$_2$Ti$_2$O$_7$, $T = 1.1$~K, 1.2~K, and 1.5~K\@. Lines: linear fits ($T_1 \propto B$) to high-field data.} 
 \end{figure}
Above a crossover field~$B_{\rm c}$ that decreases from $\sim$2.5~T at 1.5~K to $\sim$1~T at 1.1~K $T_1$ is proportional to field (fit lines in Fig.~1). Below $B_{\rm c}$ $T_1$ tends towards a constant except perhaps at 1.1~K\@. A linear field dependence of $T_1$ (with no crossover at low fields) was also found at low temperatures in the diluted pyrochlore Ising antiferromagnets~(Tb$_p$Y$_{1-p}$)$_2$Ti$_2$O$_7$, $0.21 \le p \le 1$~\cite{KGEF04}, which do not order magnetically down to $\sim$50 mK. Such behavior has been called ``cooperative paramagnetism'', since the interactions that couple the paramagnetic spins strongly affect the spin dynamics but do not cause the spins to order. Our measurements suggest that similar cooperative paramagnetism characterizes the high-temperature phase of Gd$_2$Ti$_2$O$_7$.

In general the muon relaxation rate~$1/T_1$ in a paramagnet is given by~\cite{Mori56}
\begin{equation}
1/T_1 \propto (T/\omega_\mu)\sum_{\bf q} |H_{\rm hf}({\bf q})|^2 \chi''({\bf q},\omega_\mu) \,,
\label{eq:Moriya}
\end{equation}
where $H_{\rm hf}({\bf q})$ is the spatial Fourier transform of the transferred hyperfine interaction between a muon and the paramagnetic spin system, $\chi''({\bf q},\omega_\mu)$ is the dissipative component of the complex spin susceptibility, and $\omega_\mu = \gamma_\mu B$ is the muon Zeeman frequency, where $\gamma_\mu = 852\ \mu\rm s^{-1}\ T^{-1}$ is the muon gyromagnetic ratio. In the sum over {\bf q} of Eq.~(\ref{eq:Moriya}) $\chi''({\bf q},\omega_\mu)$ is weighted by the hyperfine coupling factor (form factor)~$|H_{\rm hf}({\bf q})|^2$. If $H_{\rm hf}({\bf q})$ depends significantly on {\bf q}, the effective weighting changes as $\chi''({\bf q},\omega_\mu)$ evolves from a {\bf q}-independent form at high temperatures (appropriate to independently-fluctuating local moments) to a form with a peak at the ordering wave vector as the transition is approached. The physical picture for this behavior is that the fluctuating local field at a given muon site is the resultant of transferred hyperfine fields from a number of neighboring spins, and will therefore be modified by any change in spatial correlation between the fluctuations of these neighboring spins.

If either $|H_{\rm hf}({\bf q})|^2$ or $\chi''({\bf q},\omega_\mu)$ is substantially independent of {\bf q}, however, the temperature and field dependencies of $1/T_1$ reflect those of the local dissipative susceptbility~$\chi''(\omega) = \sum_{\bf q} \chi''({\bf q},\omega)$. The {\bf q} dependence of $H_{\rm hf}$ can be estimated qualitatively by comparing the paramagnetic-state hyperfine field~$A_{\rm hf}$, which gives an estimate of the effective coupling for uncorrelated spins, with the zero-field $\mu$SR frequency or frequencies observed in the magnetically ordered state, which yield the effective coupling for the ordered-state spin configuration.

In Gd$_2$Ti$_2$O$_7$ $A_{\rm hf} = 18.9(2)\ {\rm mT}/\mu_{\rm B}$ from trans\-verse-field $\mu$SR experiments at high temperatures~\cite{DKCC03}. In the magnetically ordered state, where Gd moments are oriented parallel to local $\langle 111\rangle$ directions~\cite{BHOS03}, two muon frequencies, $21.2(1)\ {\rm mT}/\mu_{\rm B}$ and $27.7(1)\ {\rm mT}/\mu_{\rm B}$ (assuming an ordered Gd$^{3+}$ moment of $7\ \mu_{\rm B}$), are observed in zero applied field~\cite{YDGM05}. The coupling strengths in the paramagnetic and magnetically ordered states do not differ by orders of magnitude. This indicates that $H_{\rm hf}({\bf q})$ is roughly independent of {\bf q}, and suggests that the temperature and field dependencies of $1/T_1$ are roughly those of $\chi''(\omega_\mu)/\omega_\mu$ [cf.\ Eq.~(\ref{eq:Moriya})]. The weak field dependence of $T_{\rm m}$ noted above also suggests that the form of $\chi''({\bf q},\omega_\mu)$ is substantially independent of field, as assumed in our analysis. 

The fluctuation-dissipation theorem relates $\chi''(\omega)$ to the temporal Fourier transform of the local $f$-electron spin autocorrelation function~$\langle {\bf S}(t)\,{\boldsymbol{\cdot}}\,{\bf S}(0)\rangle$ $[{\bf S}(t) = \sum_{\bf q} {\bf S}({\bf q},t)]$. For a power-law form~$\langle {\bf S}(t)\,{\boldsymbol{\cdot}}\,{\bf S}(0)\rangle \propto t^{-x}$, expected under some conditions for critical fluctuations near a thermodynamic or quantum phase transition, this yields $T_1 \propto B^{1-x}$ as the muon Zeeman frequency~$\gamma_\mu B$ is swept through the singular spin-fluctuation noise spectrum~\cite{KGEF04,KMCL96,KBCL01}. For $x \ll 1$~\cite{KGEF04}
\begin{equation}
T_1 = (1/\pi\gamma_\mu H_\mu^2 x)\,B \,,
\end{equation}
where $H_\mu$ is the rms fluctuating field at the muon site. The slope~$dT_1/dB \approx 5\ \mu\rm s/T$ in Tb$_2$Ti$_2$O$_7$~\cite{KGEF04}, whereas from the data of Fig.~1 we find a much smaller value, $\approx 0.16~\mu\rm s/T$, in Gd$_2$Ti$_2$O$_7$ close to $T_{\rm m}$. This may reflect a larger value of $x$ and/or a larger coupling field~$H_\mu$ in Gd$_2$Ti$_2$O$_7$. The crossover to $T_1 \approx \rm const.$ for $B \lesssim B_{\rm c}$ suggests that there is a cutoff of $\langle {\bf S}(t)\,{\boldsymbol{\cdot}}\,{\bf S}(0)\rangle$ at long times~$\gtrsim \tau_{\rm c} = (\gamma_\mu B_{\rm c})^{-1} \approx 10^{-9}$~s. That is, the time dependence of $\langle {\bf S}(t)\,{\boldsymbol{\cdot}}\,{\bf S}(0)\rangle$ has the more general form~$t^{-x}f(t/\tau_{\rm c})$ expected near a phase transition~\cite{KBCL01}, with a cutoff time~$\tau_{\rm c}$ two orders of magnitude longer than the time scale~$\hbar/k_{\rm B}T_{\rm m} \approx 10^{-11}$~s associated with the transition.

Figure~2 shows the temperature dependence of $1/T_1$ at several applied fields. 
 \begin{figure}[ht]
     \centering
     \includegraphics[width=0.85\columnwidth]{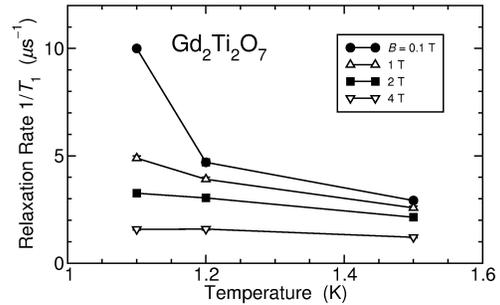}
     \caption{Temperature dependence of the longitudinal-field muon spin relaxation rate in Gd$_2$Ti$_2$O$_7$, applied field~$B = 0.1$~T, 1~T, 2~T, and 4~T (data from Fig.~1).} 
 \end{figure}  
At low fields $1/T_1$ increases markedly as $T_{\rm m}$ is approached from above. Similar behavior is generally observed in the neighborhood of a magnetic transition, and is interpreted as evidence for critical slowing down of the spin fluctuations (i.e., divergence of $\tau_{\rm c}$). In Gd$_2$Ti$_2$O$_7$, however, this rate increase is suppressed by applied field. At first sight this is surprising, because the high- and low-temperature phases are separated by a line of second-order phase transitions~\cite{RDGM99,RSHK02,PLBP04}, and critical slowing down would be expected at any applied field as the transition line is approached. The apparent suppression of critical slowing down is, however, consistent with the cutoff power-law divergence scenario outlined above: for $B \gtrsim 1/\gamma_\mu \tau_{\rm c}$ the power-law behavior of $\langle {\bf S}(t)\,{\boldsymbol{\cdot}}\,{\bf S}(0)\rangle$, not the divergence of $\tau_{\rm c}$, controls $1/T_1$. Critical slowing down would not be suppressed in systems where $\tau_{\rm c}$ is sufficiently small, i.e., $\tau_{\rm c} \ll 1/\gamma_\mu B$ for measurements at applied field~$B$ (cf.\ Fig.~2).

\section{Conclusions}
Our results suggest that spin fluctuations above the magnetic phase transition in Gd$_2$Ti$_2$O$_7$ are associated with cooperative paramagnetic behavior similar to that found in (Tb$_p$Y$_{1-p}$)$_2$Ti$_2$O$_7$~\cite{KGEF04}, with the difference that a long-time cutoff causes saturation of the fluctuation spectrum at low frequencies (low applied fields). It is the anomalously low frequency of this crossover that suppresses the critical slowing down and permits observation of the high-field power-law field dependence. To our knowledge this behavior has not been anticipated theoretically.

\section*{Acknowledgment}
We are grateful to S. Kreitzman, B. Hitti, D. Arsenau, and C. Kaiser for their help during the experiments, and to K. Ishida and S. Fujimoto for useful discussions. One of us (D.E.M.) wishes to thank Y. Maeno and the Quantum Materials Group at Kyoto University, and also K. Nagamine and the MSL group at KEK, Tsukuba, for their hospitality during visits when part of this work was carried out. This research was supported by U.S. NSF Grants DMR-0102293 (Riverside) and DMR-0203524 (Los Angeles). Work at Los Alamos was carried out under the auspices of the U.S. DOE.

\end{document}